\documentclass[3p, times, twocolumn, sort&compress]{elsarticle}
\usepackage{color}
\usepackage{amssymb}
\usepackage[german,english]{babel}
\usepackage{amsfonts}
\usepackage{amsmath,bm}
\usepackage{array}
\usepackage{epsf}
\usepackage{epsfig}
\usepackage{float}
\usepackage[version=3]{mhchem}
\usepackage{textcomp}
\usepackage{graphics}
\usepackage{graphicx}
\usepackage{epstopdf}
\usepackage{placeins}
\usepackage{footnote}
\usepackage[para]{threeparttablex}
\makesavenoteenv{tabular}
\makesavenoteenv{table}
\newcolumntype{x}[1]{>{\centering\arraybackslash\hspace{0pt}}p{#1}}

\newcommand{\ok}[1]{\textcolor{black}{#1}}

\begin{document}
\def\cm{cm$^{-1}$}
\newcommand*{\eref}[1]{Eq.~(\plainref{#1})}
\newcommand*{\fref}[1]{Fig.~\plainref{#1}}
\newcommand*{\tref}[1]{Tab.~\plainref{#1}}
\newcommand{\onlinecite}[1]{\hspace{-1 ex} \nocite{#1}\citenum{#1}} 
\newcommand*{\bref}[1]{Ref.~\onlinecite{#1}}

\hyphenation{fa-shion}
\hyphenation{o-ther-wise}
\hyphenation{cha-rac-te-ri-za-tion}

\begin{frontmatter}
\title{Manipulating the Dynamics of a Fermi Resonance with Light. A Direct Optimal Control Theory Approach}
%
\author{Alejandro R. Ramos Ramos}
\author{Oliver K\"uhn}
\ead{oliver.kuehn@uni-rostock.de}
\cortext[cor1]{Corresponding author}
\address
{Institut f\"ur Physik, Universit\"at Rostock, Albert-Einstein-Str. 23-24, D-18059 Rostock, Germany}

\begin{abstract}
Direct optimal control theory for quantum dynamical  problems presents itself as an interesting alternative to the traditional indirect optimal control. The   method relies on the first discretize and then optimize paradigm, where a discretization of the dynamical equations leads to a nonlinear optimization problem. It has been applied successfully to the control of a bistable system where the wavepacket had been approximated by a parameterized Gaussian, leading to a semiclassical set of equations of motion (A. R. Ramos Ramos, O. K\"uhn, Front. Phys. 9 (2021) 615168). Motivated by these results, in the present paper we extend the application of the method to the case of exact wavepacket propagation using the example of a generic Fermi-resonance model. In particular we address the question how population of the involved overtone state can be avoided such as to reduce the effect of intramolecular vibrational energy redistribution. A methodological advantage is that direct optimal control theory  offers  flexibility when choosing the running cost, since there is no need to compute functional derivatives and coupling terms as in the case of indirect optimal control. We exploit this fact to include state populations in the running cost, which allows their optimization.
\end{abstract}
\end{frontmatter}
\cleardoublepage

\section{Introduction}
Controlling molecular quantum dynamics with laser light has become a mature research area~\cite{keefer18_2279}. \ok{On the experimental side feedback control has been very successful~\cite{judson92_1500,assion98_5,brixner03_418}, for simulations  optimal control theory (OCT) is the workhorse for about three decades}  \cite{peirce88_4950,kosloff89_201,brif10_075008,worth13_113,keefer18_2279}. Traditionally one distinguishes global and local optimal control, depending on whether 
information on the whole time interval is included into the pulse determination or not \cite{engel09_29}.
The original OCT has been extended into several directions to take into account, for instance, constraints on the laser pulse spectra such as to accommodate experimental conditions~\cite{gollub08_073002} or to trigger quantum logic operations \cite{devivie07_5082}. The latter is an example of so-called multi-target optimization, which had been developed in Ref.\ \cite{ohtsuki01_8867} in the context of control of intramolecular vibrational redistribution (IVR). In  Ref. \citenum{ohtsuki01_8867} IVR was exploited to the advantage of the control aim. In general, however, IVR is considered being a factor limiting control yields \cite{zhao00_4882,tremblay08_194709} Unwanted loss channels due to IVR and decoherence or, in the strong field case, (multiple) ionization have been the driving force behind the development of methods incorporating state-dependent constraints. Here, an unwanted subspace is identified whose population shall be kept at minimum. Palao et al.  \cite{palao08_063412} developed a monotonically convergent Krotov type method for a functional which integrates the expectation value of projector onto an unwanted subspace over the propagation time interval. The performance was demonstrated for Rb$_2$ including three potential energy curves. Transitions between two electronic states were driven while  suppressing the population of the third state. Later applications included the control of 
coherence in a six-level $\Lambda$-system, avoiding population of an intermediate and certain spectator states to potentially enhance a coherent anti-Stokes Raman scattering (CARS) signal~\cite{kumar13_326} or the control of rotation of a linear molecule by excluding certain rotational levels such as to tune the time for free rotation~\cite{ndong14_857}. \ok{ The respective method had been proposed in Ref.\ \citenum{lapert12_033406}  as a means to minimize relaxation effects. It should be noted that these global OCT formulations had  a predecessor in local optimal control, i.e. the so-called optical paralysis discussed by Malinovsky and coworkers in the 1990s. These authors could show that vibrational dynamics in  the electronic ground state can be triggered via  impulsive stimulated Raman scattering without population of the electronically excited state. This can be achieved solely by manipulating the phase of the field \cite{malinovsky97_67}.}

Global OCT is an \textit{indirect} method which follows the first optimize and then discretize paradigm. Here the condition of  optimality on the performance functional leads to a set of coupled Schr\"odinger equations that are solved using iterative forward-backward sweep methods. From all explored algorithms to find the stationary solution of the coupled Schr\"odinger equations, the Krotov method is most commonly used because it guarantees monotonic convergence \cite{reich_2012_Monotonically}. 
By construction any change in the performance functional leads to a different set of working equations and thus a modification of the Krotov algorithm is needed, as shown, e.g., for time-dependent targets \cite{serban_2005_Optimal}, unitary transformations \cite{palao_2003_Optimal}, constrained laser spectra \cite{gollub08_073002}, state dependent constraints \cite{palao08_063412} \ok{or final time optimization \cite{ndong14_857}}.

\textit{Direct} optimal control, in contrast, relies on the first discretize and then optimize paradigm. It converts the initial optimal control problem into a nonlinear optimization problem after direct discretization of the performance functional. Direct optimal control comes with the advantage that there is no need to analytically solve for the  stationarity condition of the performance functional. Thus,  a wide variety of terms can be included in the performance functional in a plug-and-play fashion without tedious derivations.  Additionally, otherwise fixed quantities like the final time and the initial state could be also subject to optimization. Although being popular, for instance, in applied mathematics~\cite{betts10_}, engineering~\cite{pardo16_946}, and biology~\cite{chen-charpentier20_112983}, an  application to quantum molecular dynamics has been presented only recently in Ref. \citenum{ramos21_615168}. In that paper we have provided a proof-of-principle study focusing on the control of a wavepacket in a bistable potential. Thereby, the quantum dynamics has been approximated by propagation of a single Gaussian wavepacket.

The goal of the present paper is to provide a first application of direct optimal control to trigger quantum dynamics beyond the Gaussian wavepacket approximation with state dependent constraints. Specifically, we will focus on the situation of a  Fermi resonance  between the fundamental transition of a stretching vibration and the first overtone transition of the associated bending vibration. Fermi-resonance interaction in hydrogen bonds is known for its effect on the absorption lineshape~\cite{henri-rousseau02_241}. It also provides the pathway for rapid IVR in hydrogen bonds upon excitation of the hydrogen stretching vibration~\cite{kuhn02_7671,heyne04_6083}. The questions to be addressed here are whether (i) a stretching vibration can be prepared without simultaneous bending overtone excitation and (ii) if the stretching excitation once prepared can be effectively decoupled from the bending overtone such as to reduce Fermi-resonance mediated IVR. \ok{These questions can be put into the more general context of manipulating zero-order states with taylored laser fields \cite{abdel-latif_2011_Carbonyl,lisaj_2014_Laserdriven,santos18_2213}.
}

The paper is organized as follows: Section 2 will briefly summarize the theoretical framework of direct optimal control theory; the  Fermi-resonant model will be introduced in Section 3; in Section 4 we will focus on different control scenarios chosen such as to illustrate the capabilities of the approach and finally, a summary  will be given in Section 5.

\section{Direct Optimal Control Theory (DOCT)}
\label{sec:DOCT}
In the following we will apply DOCT to a system with Hamiltonian $\hat{H}_{0}$  coupled to an electric field within dipole approximation, $\hat{H}_{\rm F}(t)$. The total Hamiltonian reads in the Schr\"odinger picture
\begin{equation}
	\hat{H}(t)=\hat{H}_{0} + \hat{H}_{\rm F}(t)~,
\end{equation}
with $\hat{H}_{\rm F}(t)=-\hat{\mu} E(t)$. Note that we will not consider effects due to the vector character of dipole moment, $\hat{\mu}$, and field, $E(t)$. For the time-dependent Schr\"odinger equation it is convenient to switch to the interaction representation with respect to $\hat{H}_{0}$, i.e. $|\Psi^{\rm (I)}(t)\rangle = \exp(i\hat H_0t/\hbar) |\Psi(t)\rangle$, in order to obtain a smoother dynamics, 
\begin{equation}
\label{eq:TDSE}
i\hbar \frac{\partial}{\partial t} | \Psi^{(\rm I)}(t) \rangle = \hat{H}^{\rm (I)}_{\rm F}(t) | \Psi^{\rm(I)}(t) \rangle\, .
\end{equation}
Suppose that the eigenvalue equation of the system has been solved, i.e. $\hat{H}_{0}|m\rangle=E_m|m\rangle$ ($m=1,\ldots,M_{\rm max}$), Eq. \eqref{eq:TDSE}, can be projected to the eigenstate $|m\rangle$ which gives the two equations
\begin{align}\label{eq:re_dyn}
\frac{\partial}{\partial t}  \Re \langle m |\Psi^{(\rm I)}(t) \rangle &= -  \frac{1}{\hbar}E(t)  \sum_n\mu_{mn} \Big[ \cos (\omega_{mn} t)   \Im \langle n | \Psi^{(\rm I)}(t) \rangle  \nonumber\\  &+   \sin (\omega_{mn} t) \Re  \langle n | \Psi^{(\rm I)}(t) \rangle  \Big]\, ,
\end{align}
\begin{align}\label{eq:im_dyn}
\frac{\partial}{\partial t}  \Im\langle m | \Psi^{(\rm I)}(t) \rangle & =  \frac{1}{\hbar}E(t) \sum_n \mu_{mn} \big[ \cos (\omega_{mn} t)   \Re \langle n | \Psi^{(\rm I)}(t) \rangle    \nonumber\\  &-   \sin (\omega_{mn} t) \Im  \langle n | \Psi^{(\rm I)}(t) \rangle  \big]\, ,
\end{align}
where $\mu_{mn}=\langle m | \hat{\mu} | n \rangle$ and $\omega_{mn}=(E_m-E_n)/\hbar$.
The system of dynamic equations \eqref{eq:re_dyn}-\eqref{eq:im_dyn} can be cast into the following form:
\begin{equation}\label{eq:diffcon}
  \dot{\bm a}(t)=f[\bm a(t),\bm u(t),\bm k,t]~,
\end{equation}
where we defined the state vector ${\bm a}(t)=(\Re\langle m | \Psi^{(\rm I)}(t) \rangle, \Im\langle m | \Psi^{(\rm I)}(t) \rangle)^{\rm T}$ with $m=1,\ldots,M_{\rm max}$. Further, $\bm u(t)=E(t)$ is the external laser field and $\bm k$ is a set of time-independent parameters.

The general control problem can be formulated as follows \cite{brif10_075008,werschnik07_R175,worth10_15570,becerra10_1391,ramos21_615168}: Given a performance functional of the form
\begin{equation}\label{eq:J}
  {\mathcal J}[\bm a, \bm u, \bm k]= {\mathcal T}[\bm a(t_{\rm f}),\bm k,t_{\rm f}] + \int_{t_0}^{t_{\rm f}} {\mathcal R}[\bm a(t),\bm u(t),\bm k,t]\,dt~,
\end{equation}
where ${\mathcal T}$ and ${\mathcal R}$ are the terminal and running cost, respectively, the task is to find on the interval $t\in [t_0,t_{\rm f}]$ the state trajectory $\bm a(t)$, external control $\bm u(t)$, and the set of static parameters $\bm k$ that minimize the functional ${\mathcal J}[\bm a, \bm u, \bm k]$.
The minimization is performed subject to the differential constraints eq. \eqref{eq:diffcon} in the interval $t\in [t_0,t_{\rm f}]$.
Further, there can be path constraints
\begin{equation}
  \bm h_{\rm L} \leq \bm h[\bm a(t),\bm u(t),\bm k,t]\leq \bm h_{\rm U}~,
\end{equation}
and event constraints such as
\begin{equation}
\bm e_{\rm L} \leq \bm e[\bm F[\bm a(t),\bm u(t)],\bm k,t_0,t_{\rm f}]\leq \bm e_{\rm U}~.
\end{equation}
Here, the subscript ${\rm L}$ and and ${\rm U}$ denotes the lower and upper boundary, respectively, defining the constraints. Notice that in contrast to path constraints, event constraints are  time-independent, but could include a functional, $\bm F$, of, e.g., the state trajectory or the external control.

For the present illustration our goal is as follows: Given some initial state $|\Psi(t_{\rm 0})\rangle=|\Phi^{\rm i}\rangle$, characterized by the parameters $\bm a^{\rm i}$, find a laser field $E(t)$ such that the overlap is maximized between the time-evolved final state at $t=t_{\rm f}$, $|\Psi(t_{\rm f})\rangle$, and some target state $|\Phi^{\rm t}\rangle$, characterized by the parameters $\bm a^{\rm t}$. Simultaneously,  minimize the energy input of the laser field and the projection onto some unwanted states during the whole time interval. 

Thus the terminal cost is given by (notice the minus sign because the performance functional will be minimized and we want to maximize the overlap)
\begin{equation}
\begin{split}
	{\mathcal T}[\bm a(t_{\rm f}),\bm k,t_{\rm f}] & = -\left| \langle  \Phi^{\rm t}  |  \Psi(t_{\rm f}) \rangle \right|^2~.
\end{split}
\end{equation}
The running cost will be chosen as follows
\begin{align}\label{eq:run_cost}
{\mathcal R}[\bm a(t),\bm u(t),\bm k,t] &={\mathcal R}_1[\bm u(t),t] + {\mathcal R}_2[\bm a(t)]\\
{\mathcal R}_1[\bm u(t),t] &=\kappa \frac{ |E(t)|^2}{s(t)} \\
{\mathcal R}_2[\bm a(t)] & = \langle  \Psi(t)|  \hat{\Pi}  |   \Psi(t) \rangle \, .
\end{align}
\ok{ The first term, ${\mathcal R}_1[E(t),t]$, provides a penalty for the field intensity scaled by the parameter $\kappa$: smaller values of $\kappa$ allow for larger field intensities and vice versa.} Further, we have included a shape function $s(t)=\sin^2 \left([ \pi/(t_{\rm f}-t_{\rm 0})] (t-t_{\rm 0})\right) +\epsilon$, which  ensures that the field increases(decreases) slowly when turned on(off) \cite{sundermann00_1896}. Note that $\epsilon$ is a small parameter introduced to avoid division by zero and numerical problems at times $t=t_{\rm 0}$ and $t=t_{\rm f}$. For the applications below we have used $\epsilon=0.005$. 

The second term, ${\mathcal R}_2[\bm a(t)]$, in the running cost defines penalties for populations of certain quantum states $|\phi_i\rangle$ via the operator
\begin{equation}\label{eq:projector}
  \hat{\Pi}=\sum_{i=1}^{n_{\mathcal R}} \kappa_i | \phi_i \rangle \langle \phi_i |~,
\end{equation}
whose expectation value is minimized over the complete trajectory. The set of parameters $\kappa_i$ scales the penalty for the population of each state $| \phi_i \rangle$ individually and they could be positive or negative, depending if we want to minimize or maximize respectively the population of $| \phi_i \rangle$. It is important to note that this contrasts with the traditional approach in multi-target OCT where the populations of the states that should be minimized are only included in the terminal cost~\cite{ohtsuki01_8867}. 

According to  Refs.~\citenum{ramos21_615168} and \citenum{becerra10_1391}, there could be path and event constraints. For the application presented below we don't use any path constraints, but event constraints. Given the event
\begin{equation}\label{event}
\bm e[\bm a(t_0),F[E(t)]]=
  \begin{pmatrix}
    \bm a(t_0) \\
    \int_{t_0}^{t_{\rm f}} E(t) dt
  \end{pmatrix}~,
\end{equation}
upper and lower bounds will be chosen equal as follows
\begin{equation}
  \bm e_{\rm L}= \bm e_{\rm U}=
\begin{pmatrix}
  \bm a^{\rm i}\\
  0
  \end{pmatrix}~.
\end{equation}
Hence, the parameters of the initial state are fixed and not subject to optimization. Further, we enforce the zero-net-force condition by demanding that $F[E(t)]= \int_{t_0}^{t_{\rm f}} E(t) dt=0$~\cite{doslic06_013402}.

The optimization problem will be solved using a direct method, i.e. by means of discretization of the differential equations. The PSOPT package will be used for this task~\cite{becerra10_1391}. Specifically,  we have used trapezoidal discretization with 800 nodes, which offers a good balance between accuracy and speed for the present case; for details on the numerical performance see Ref. \citenum{ramos21_615168}. 
\section{Model System} 
\label{sec:fermi_res}
We consider the basic model for a  Fermi-resonance interaction between fundamental and first overtone transitions. Here, two harmonic modes, $q_1$ and $q_2$, are coupled by a third-order anharmonicity. The system Hamiltonian can be written as
\begin{equation}
\label{eq:h0}
  \hat{H}_{0}=\sum_{i=1}^2 \frac{\hbar \omega_i}{2} \Bigg(  - \frac{\partial^2}{\partial q_i^2} + q_i^2 \Bigg)+ \gamma \hbar \omega_2 q_1 q_2^2 ~.
  \end{equation}
  Here, $q_i$ are the dimensionless oscillator coordinates (length scale for mass-weighted coordinates $\lambda_i=\sqrt{\hbar /\omega_i}$) and we assumed for simplicity an exact resonance between the fundamental transition of $q_1$ and the first overtone of $q_2$, i.e. $\omega_1=2\omega_2$. Below, all energies will be given in units of $\hbar\omega_2$. The parameter $\gamma$ is the dimensionless coupling strength for which a value of $\gamma = 0.07$ has been chosen to provide a moderate coupling
For the dipole moment a  linear model with equal magnitudes was assumed, i.e. $\hat \mu(q_1,q_2)=\mu_0(q_1/\sqrt{2}+q_2)$, where the factor $1/\sqrt{2}$ is due to the scaling to dimensionless coordinates.
\begin{figure}[th]
\includegraphics[width=1.0\columnwidth]{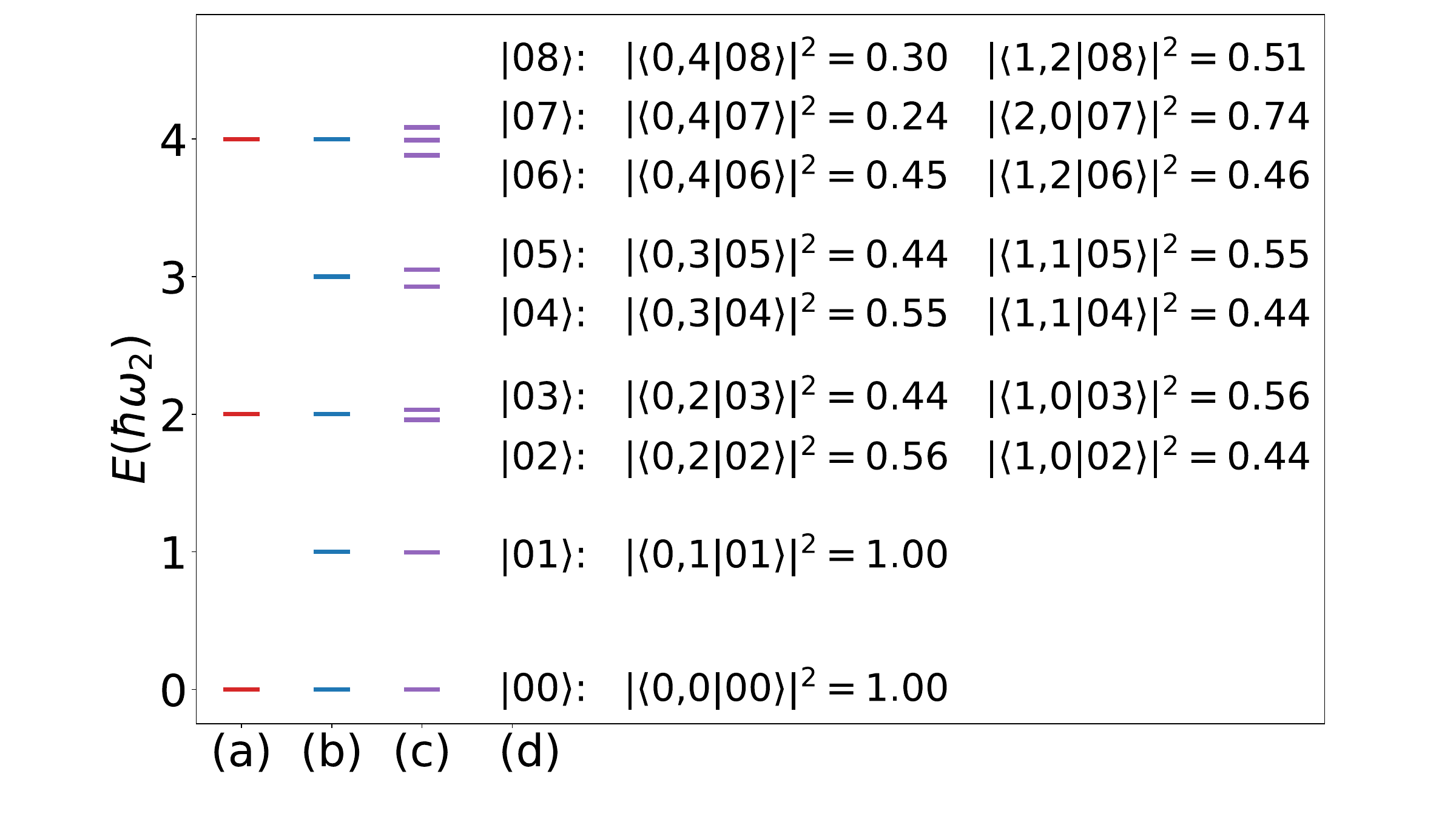}
        \caption{Harmonic oscillator energies corresponding to (a) $q_1$, (b) $q_2$, and (c) eigenenergies of the anharmonic model.  Panel  (d) contains a decomposition of the eigenstates into the harmonic oscillator basis (cf. Eq. \eqref{eq:eigenstates}).}\label{fig:eigenenergies}
\end{figure}

The  product basis $|n_1,n_2\rangle$ formed by the harmonic oscillator states (zero-order states) can be used to expand the eigenstates of Eq. \eqref{eq:h0} in a truncated Hilbert space of dimension $(N_1+1)\times (N_2+1)$
\begin{align}
\label{eq:eigenstates}
    |m\rangle = \sum_{n_1=0}^{N_1}\sum_{n_2=0}^{N_2} C_{n_1,n_2}(m) \,|n_1,n_2\rangle\; .
\end{align}
The corresponding eigenenergies will be labeled $E_m$. In the numerical implementation we have chosen $N_1=9$ and $N_2=19$  giving a total of 200 eigenstates after diagonalization, from which we will retain the lowest 18 for the dynamics simulations \ok{in sections \ref{sec:control_over} and \ref{sec:example2}, and 37 for the ones present in section \ref{sec:example3}. Higher lying states turned out to be marginally populated as has been checked by test calculations using the optimized field and a larger basis (up to 60 states). The differences for the cases shown below have been a few percent at most. Energies and characterization of eigenstates in terms of the harmonic oscillator basis are given in Fig. \ref{fig:eigenenergies}.} 

The dipole matrix follows accordingly, in particular we have for the transition from the ground to the Fermi-resonance pair states ($m=2,3$)
$\langle m |\hat \mu|0\rangle \propto C_{1,0}(m) \langle 1,0 | q_1  |0,0\rangle$, i.e. due to the linearity of the dipole operator only the fundamental transition of the high-frequency oscillator contributes oscillator strength (about 50\% each for $m=2$ and $3$) 
The fundamental transition of the lower-frequency mode is a pure excitation according to Fig. \ref{fig:eigenenergies}. On the other hand, the overtone transition into the Fermi-resonance has a dipole matrix element 
$\langle m |\hat \mu|0,1\rangle \propto C_{1,0}(m) \langle 1,0 | (q_1/\sqrt{2} +q_2)   |0,1\rangle + C_{0,2}(m) \langle 0,2 | q_2  |0,1\rangle$. The first term describes a simultaneous deexcitation of $q_2$ and an excitation of $q_1$ and the second term the excitation of the overtone of $q_2$. Below we will refer to this mechanism as the \textit{indirect and direct pathway}, respectively.

The optimal laser field will be analyzed employing a time window with shape
\begin{equation}
  G(t,t')=
  \begin{cases}
    1  & |t-t'|\leq \frac{w}{2}\\
    \exp\left[-  \frac{(|t-t'| - \frac{w}{2})^2}{2\sigma^2}\right]   & |t-t'|> \frac{w}{2}~.
  \end{cases}
\end{equation}
The parameter $w$, chosen as $w=17 ~\omega_2^{-1}$, describes the temporal width of the window, whereas $\sigma=2 ~\omega_2^{-1}$ gives its decay, and $t'$ is the center of the window, which has been scanned from $t_0+w/2$ to $t_{\rm f}-w/2$.

\begin{figure}[tb]
\center
        \includegraphics[width=0.8\columnwidth]{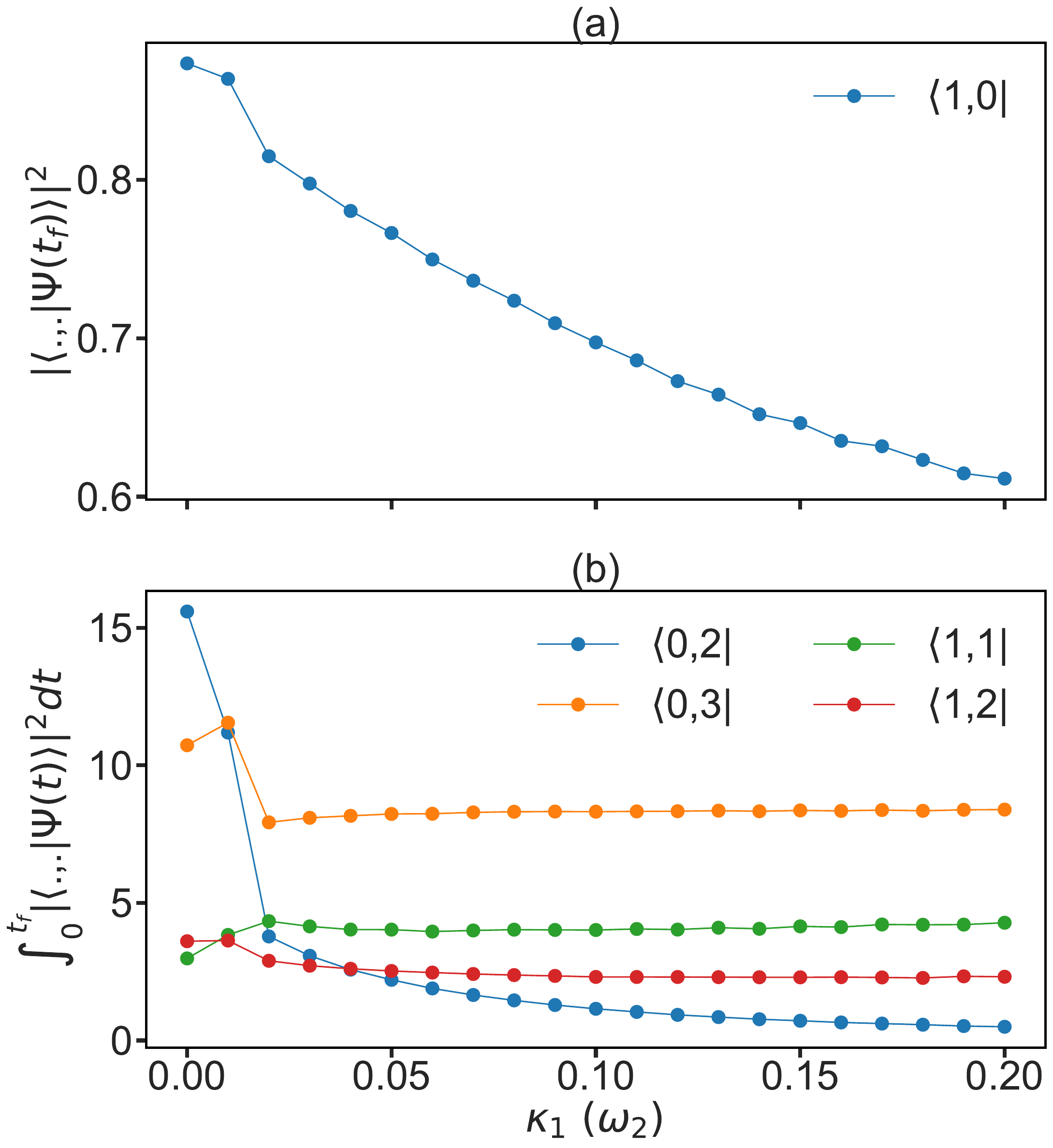}
        \caption{State-selective excitation of a Fermi-resonance pair, dependence on the penalty parameter $\kappa_1$: (a) final population of state $|1,0\rangle$, (b) integrated population of states $|0,2\rangle$, $|0,3\rangle$, $|1,1\rangle$ and $|1,2\rangle$.}\label{fig2}
\end{figure}
\begin{figure}[tb]
\center
        \includegraphics[width=0.8\columnwidth]{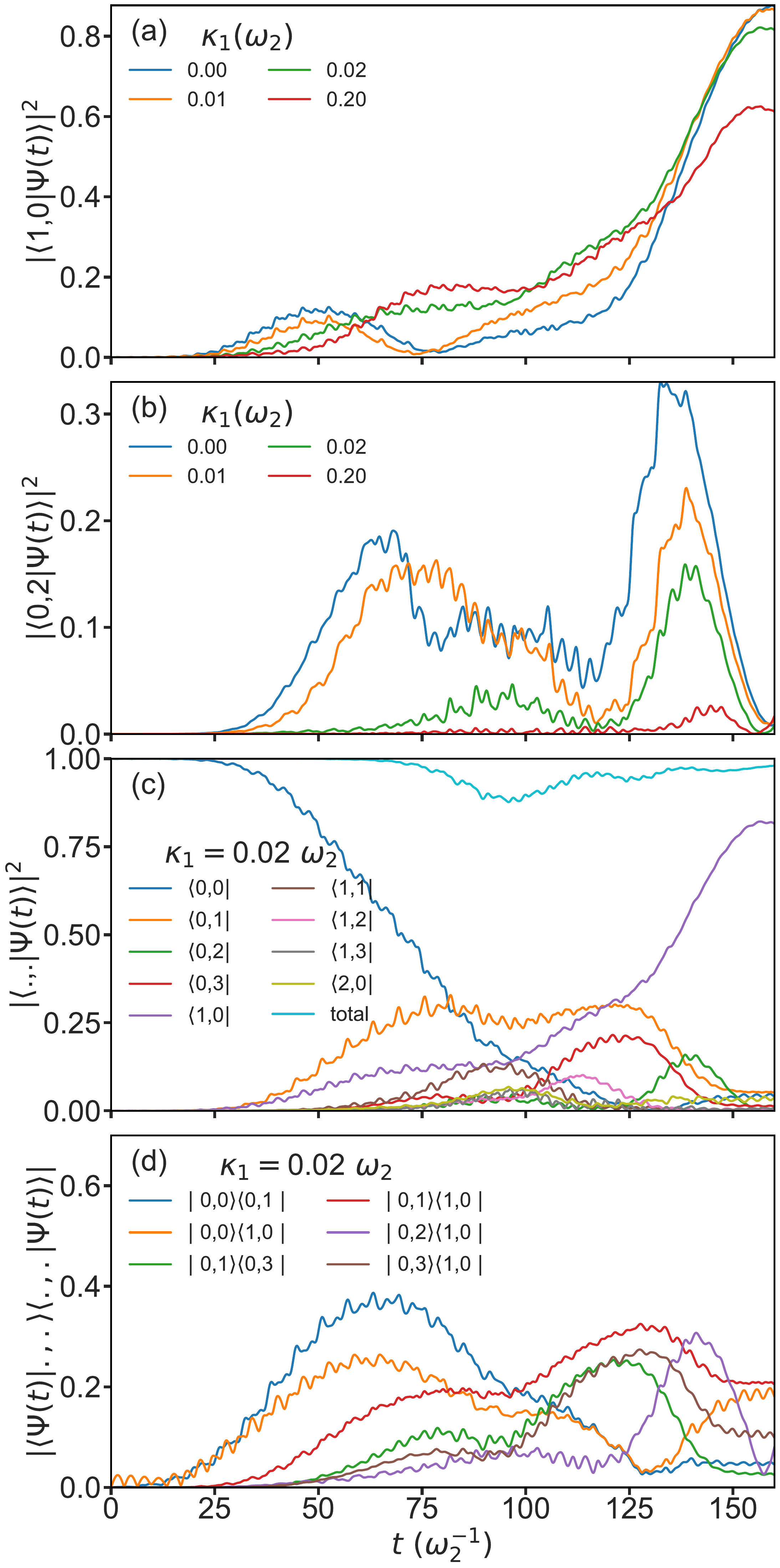}
        \caption{State-selective excitation of a Fermi-resonance pair: Population dynamics of state (a) $|1,0\rangle$ and (b) $|0,2\rangle$ for different values of $\kappa_1$. For $\kappa_1=0.02 ~\omega_2$ the population of the most relevant states are shown in (c) and the most relevant coherences are shown in (d).  }\label{fig3}
\end{figure}

\begin{figure*}[tb]
\centering
        \includegraphics[width=1.8\columnwidth]{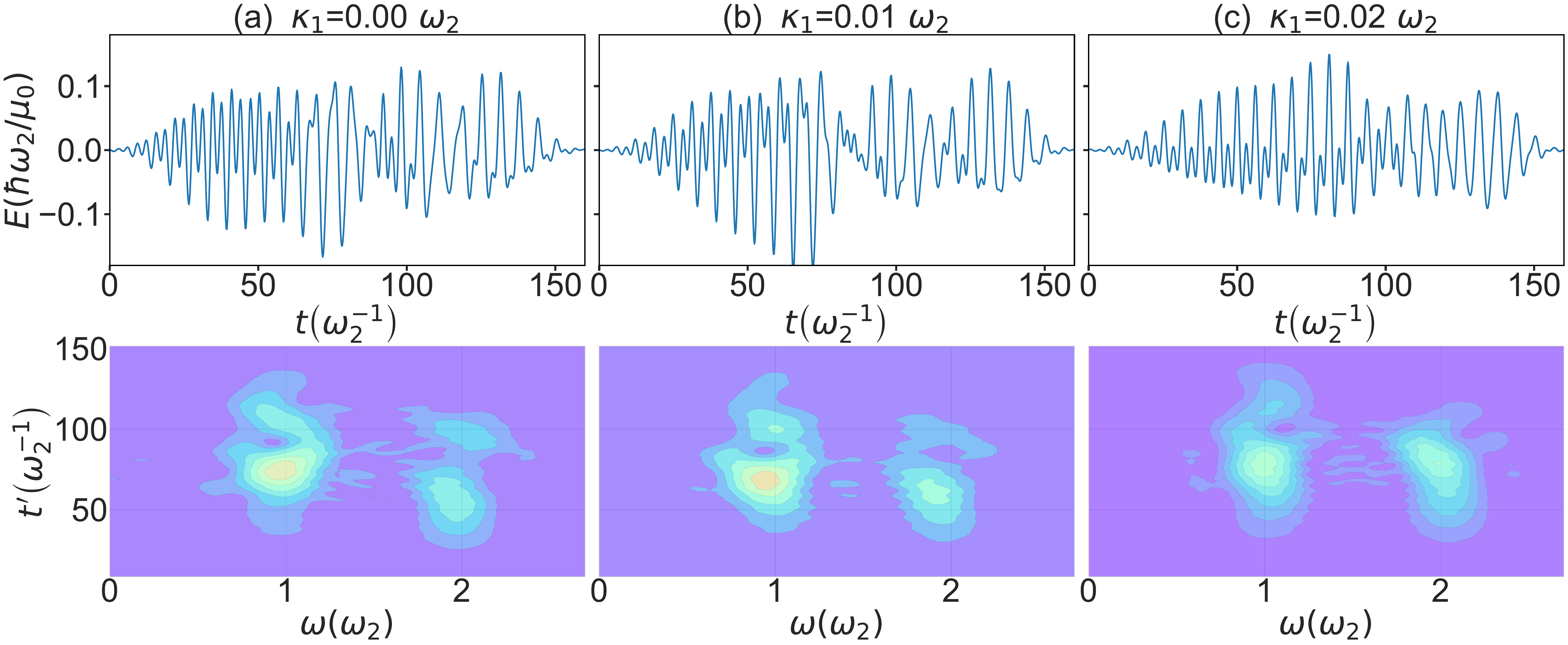}
        \caption{State-selective excitation of a Fermi-resonance pair: Optimal field (first row) and time dependent power spectrum (second row) for different values of $\kappa_1$.}\label{fig4}
\end{figure*}

\section{Results}
The anharmonic coupling between the zero-order states $|1,0 \rangle$ and $|0,2 \rangle$ has the effect that an excitation of state $|1,0 \rangle$ provokes a population transfer to state $|0,2 \rangle$, and vice versa. The state $|0,2 \rangle$ in turn could provide a gateway for rapid IVR. In order to illustrate the DOCT approach, in the following three scenarios will be discussed which eventually aim at preparing or maintaining a high population in state  $|1,0 \rangle$   while simultaneously having a small population in state $|0,2 \rangle$. Notice that in all examples we will focus exclusively on the given time interval, i.e. we do not consider the system dynamics beyond the final time where both states will exchange population according to the strength of the Fermi-resonance coupling. Further, we checked that for the used field strengths the effect of field-dressing is small, i.e. it will not be considered when discussing the mechanisms of laser control. For the analysis we will consider the density matrix with respect to  the zero-order states, i.e.   
$\rho_{n_1,n_2;m_1,m_2}(t)=\langle \Psi(t) |m_1,m_2 \rangle\langle n_1,n_2 |\Psi(t)\rangle $. For quantum states $|i\rangle$, changes in the population are related to coherences via $\dot \rho_{ii}=(2/\hbar) {\rm Im}\sum_j V_{ij}\rho_{ji}$ if $V_{ij}$ is the matrix element of the coupling. Likewise within lowest-order of perturbation theory one has $\dot\rho_{ij}=-i\omega_{ij}\rho_{ij}-(i/\hbar) V_{ij}(\rho_{ii}-\rho_{jj})$~\cite{may11}. Hence, changes in populations are driven by coherences and vice versa.

\subsection{State-selective excitation of Fermi-resonance pair}\label{sec:control_over}
The objective of this section is to maximize the population of state $|1,0 \rangle$ after a given final time, and at the same time minimize the population of $|0,2 \rangle$ over the complete duration of the laser pulse. This way, any IVR starting from state $|0,2 \rangle$ would be suppressed.
To accomplish this task we have set $|\Phi^{\rm i} \rangle=|0,0\rangle$, $|\Phi^{\rm t} \rangle=|1,0\rangle$, $n_{\mathcal R}=1$ and $|\phi_1\rangle=|0,2\rangle$, the initial time $t_{\rm_0}=0$, the final time $t_{\rm_f}=160~ \omega_2^{-1}$, $\kappa=0.2 \left(\mu_0/\hbar \omega_2\right)^2$ and $\kappa_1$ ranged from 0 to $0.2\,\omega_2$ in steps of $0.01\,\omega_2$. 

\begin{figure}[htbp]
\centering
        \centering
        \includegraphics[width=0.8\columnwidth]{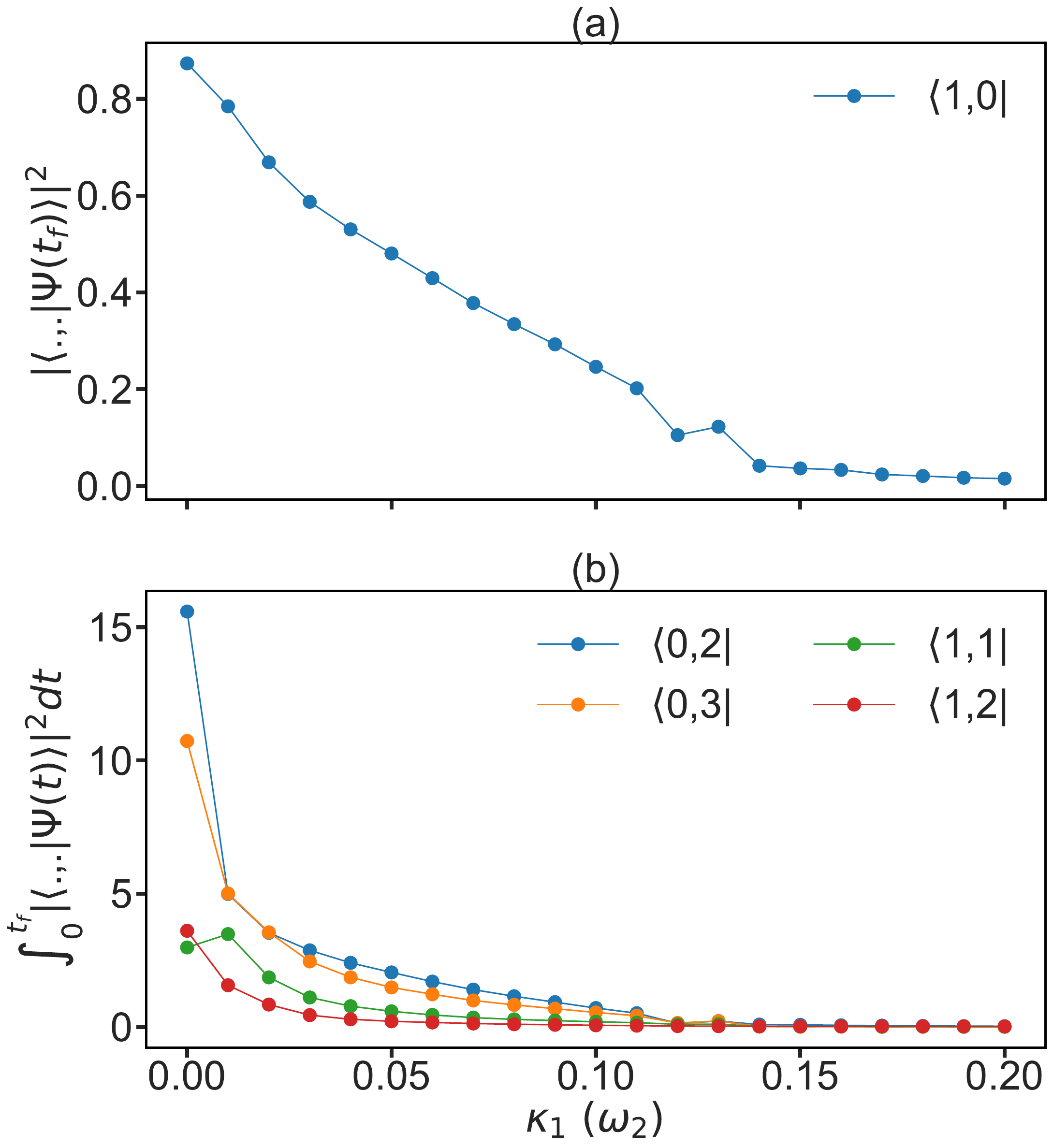}
        \caption{Suppressing overtone and combination transitions, (a) Final population of state $|1,0\rangle$, (b) integrated population of states $|0,2\rangle$, $|0,3\rangle$, $|1,1\rangle$ and $|1,2\rangle$.}\label{fig5}
\end{figure}

\begin{figure}[htbp]
        \centering
        \includegraphics[width=0.8\columnwidth]{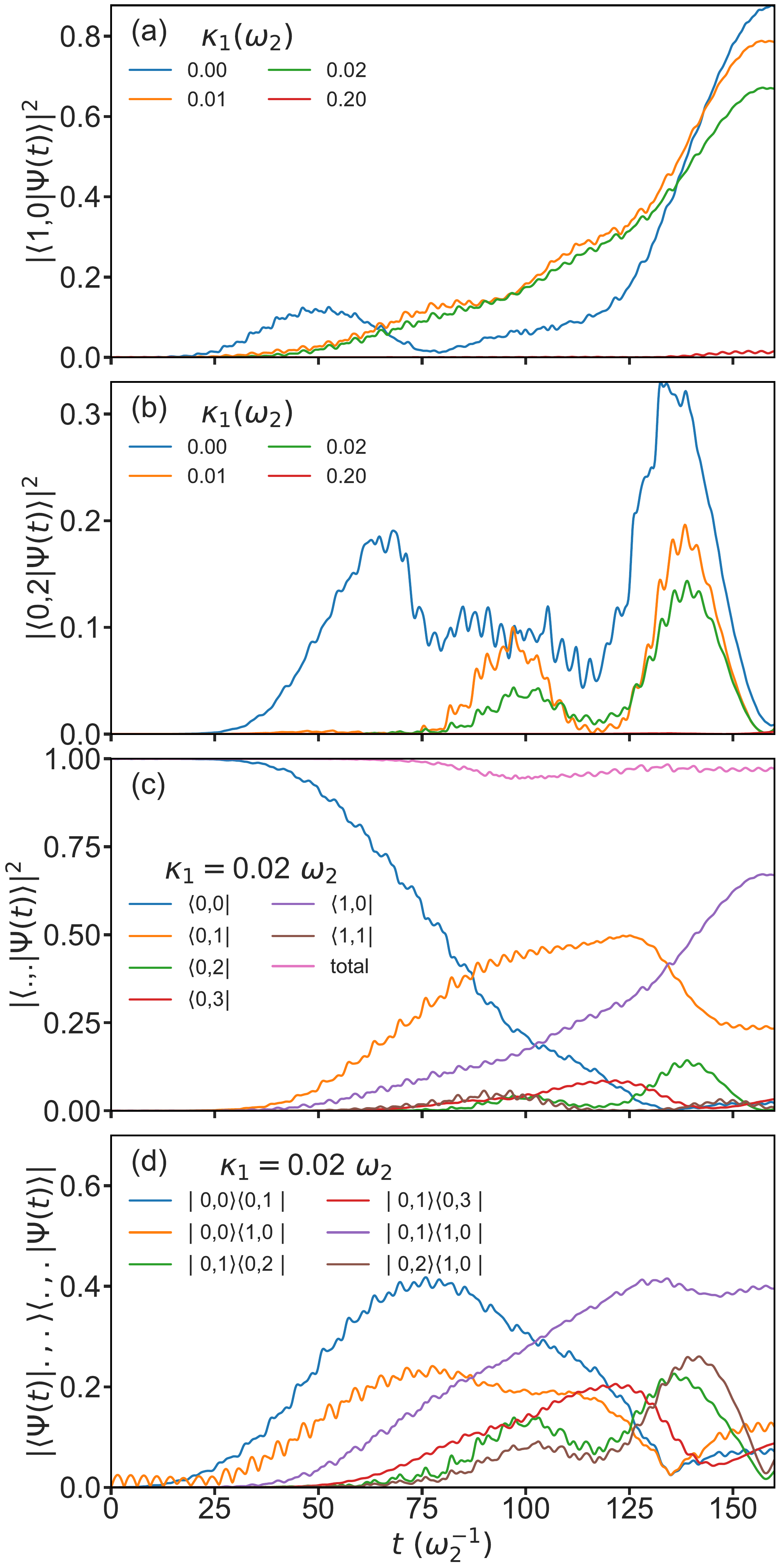}
        \caption{Suppressing overtone and combination transitions: Population dynamics of state (a) $|1,0\rangle$ and (b) $|0,2\rangle$ for different values of $\kappa_1$. For $\kappa_1=0.02 ~\omega_2$ the population of the most relevant states are shown in (c) and the most relevant coherences are shown in (d).}\label{fig6}
\end{figure}

\begin{figure*}[tbp]
        \centering
        \includegraphics[width=1.5\columnwidth]{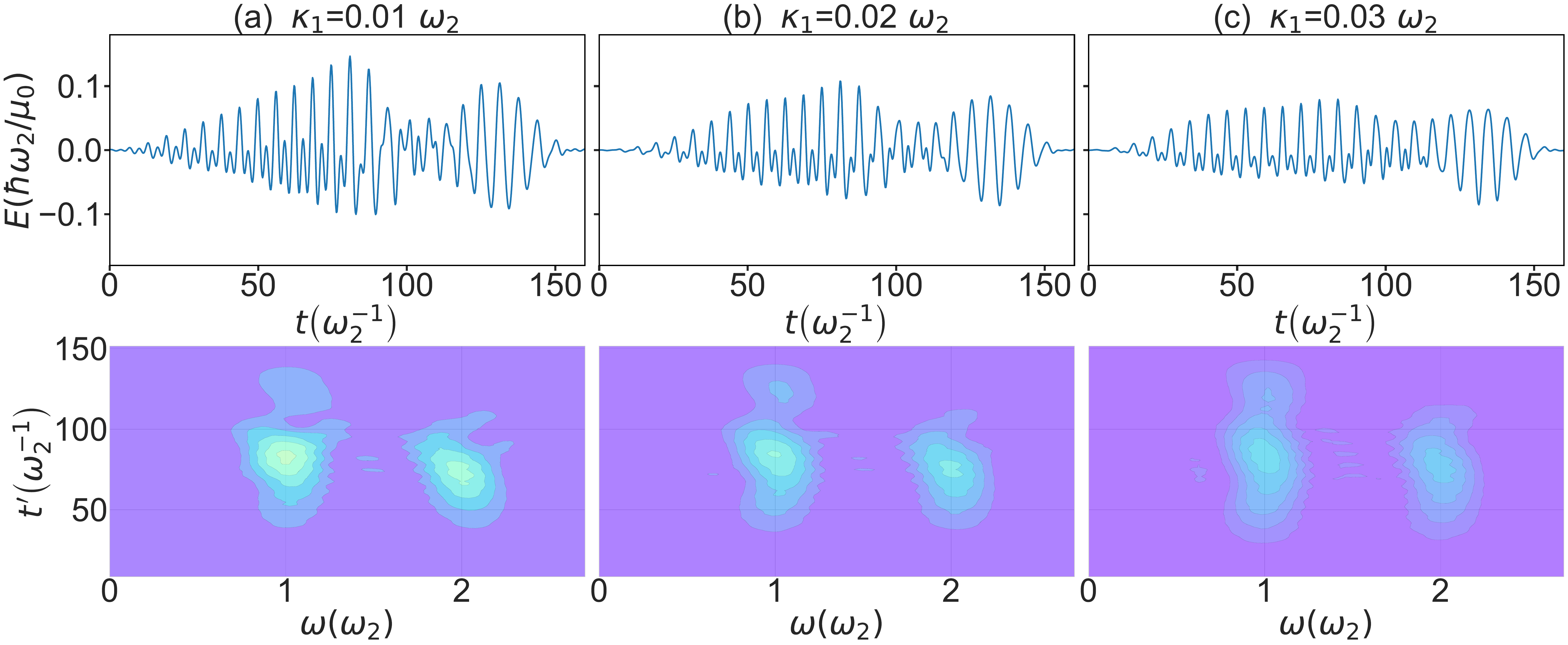}
        \caption{Suppressing overtone and combination transitions: Optimal field (first row) and time dependent power spectrum (second row) for different values of $\kappa_1$.}\label{fig7}
\end{figure*}

Results for the final population of $|1,0\rangle$ and the integral of the population of $|0,2\rangle$ are shown in Figure \ref{fig2}(a) and (b), respectively. As can be seen the integrated population of state $|0,2\rangle$ can be drastically reduced while preserving an appreciable final population of state $|1,0\rangle$.
In order to understand the laser-driven dynamics one should inspect the state populations and the coherences in Fig. \ref{fig3} together with the field analysis in Fig. \ref{fig4}.

\ok{Let us first consider the case of $\kappa_1=0$, i.e. there is no penalty for populating state $|0,2\rangle$. At the final time the target state  $|1,0\rangle$ is populated by about 85\%.} According to Figs. \ref{fig3} and \ref{fig4}(a) the pulse initially populates state $|1,0\rangle$ directly, before a ladder-climbing pathway  $|0,0\rangle \rightarrow  |0,1\rangle \rightarrow  |0,2\rangle$ sets in. \ok{The latter can be inferred from the dominance of the frequency $\omega_2$ after about 75~$\omega_2^{-1}$. } The actual laser-driven state population is superimposed by the inherent system dynamics. This can be seen by the depopulation of $|1,0\rangle$ around 75~$\omega_2^{-1}$, but even more so by the population flow between $|0,2\rangle$ and $|1,0\rangle$ toward the end of the time interval when the laser field is slowly switched-off. In other words, the control mechanism makes use of the inherent system dynamics.

Increasing $\kappa_1$ to $0.01\,\omega_2$ has little effect on the dynamics, but from about $\kappa_1=0.02\,\omega_2$ there is a qualitative change. The initial $|1,0\rangle$ excitation is suppressed, i.e. the population of this state increases more monotonously,  and the field acts with frequencies $\omega_1$ and $\omega_2$ simultaneously, see Fig. \ref{fig4}(c). According to Fig. \ref{fig3}(c) the state $|0,1\rangle$ is excited, but without appreciable ladder climbing, i.e. the population of $|0,2\rangle$ remains small until about $t\approx120\,\omega_2^{-1}$. Inspecting the coherences in  Fig. \ref{fig3}(d) one notices that part of the rise in population of $|1,0\rangle$ is due to the indirect mechanism $|0,0\rangle \rightarrow  |0,1\rangle \rightarrow  |1,0\rangle$, i.e. via the coherence density matrix element $\rho_{1,0;0,1}$. This is possible due to the dipole matrix element $\propto C_{1,0}(m) \langle 1,0 | (q_1/\sqrt{2} +q_2)   |0,1\rangle $ as discussed above. At later times $t>100\,\omega_2^{-1}$ the dynamics becomes more intricate, involving besides population of $|0,2\rangle$ also a coupling to the second overtone $|0,3\rangle$.

\subsection{Suppressing overtone and combination transitions}\label{sec:example2}
According to Fig. \ref{fig2}(b) the suppression of $|0,2\rangle$ excitation comes at the expense of having higher energetic overtone and combination transitions populated. In terms of IVR this should, of course, be avoided.
Therefore, in the following we will include penalties for the most strongly excited states, i.e $\{|\phi_i\rangle\}=(|0,2\rangle, |0,3\rangle, |1,1\rangle ,|1,2\rangle$ with $\kappa_i=\kappa_1$, which also has been ranged from 0 to $0.2\,\omega_2$ in steps of $0.01\,\omega_2$ ($n_{\mathcal R}=4$ in Eq. \eqref{eq:projector}). In Fig. \ref{fig5} the final population of $|1,0\rangle$ and the integral of the population of the states $\{|\phi_i\rangle\}$ are shown. First, one can notice that indeed excitation of these higher energetic states can be suppressed. However, it is also obvious that appreciable population of the zero-order state $|1,0\rangle$ cannot be achieved without having some population in the set $\{|\phi_i\rangle\}$. For large penalties, both the integral populations and the target state population converge to zero, with the field being zero as well.  

Figures \ref{fig6}  and \ref{fig7} show state populations for particular values of $\kappa_1$ and an analysis of the control field, respectively. Let us focus on the case of $\kappa_1=0.02\, \omega_2$. In terms of the driving field, comparing  Figs. \ref{fig4}(b) and \ref{fig7}(a), one notices that while still both frequencies are present, their amplitudes are about equal in \ref{fig7}(a) and the $\omega_2$ contribution is less structured. The population of the target state shown in Fig. \ref{fig6}(a) shows a rather monotonous rise whereas the states with a penalty expectedly have little population, Fig. \ref{fig6}(b,c). Instead the state $|0,1\rangle$ is appreciably populated. Inspecting Fig. \ref{fig6}(d) we conclude that there is a direct population of the target state and the state $|0,1\rangle$ according to the field component at $\omega_1$ and $\omega_2$, respectively. The large coherence $\rho_{1,0;0,1}$ signifies that again the indirect pathway plays an important role.

\begin{figure}[hb]
\centering
        \centering
        \includegraphics[width=0.8\columnwidth]{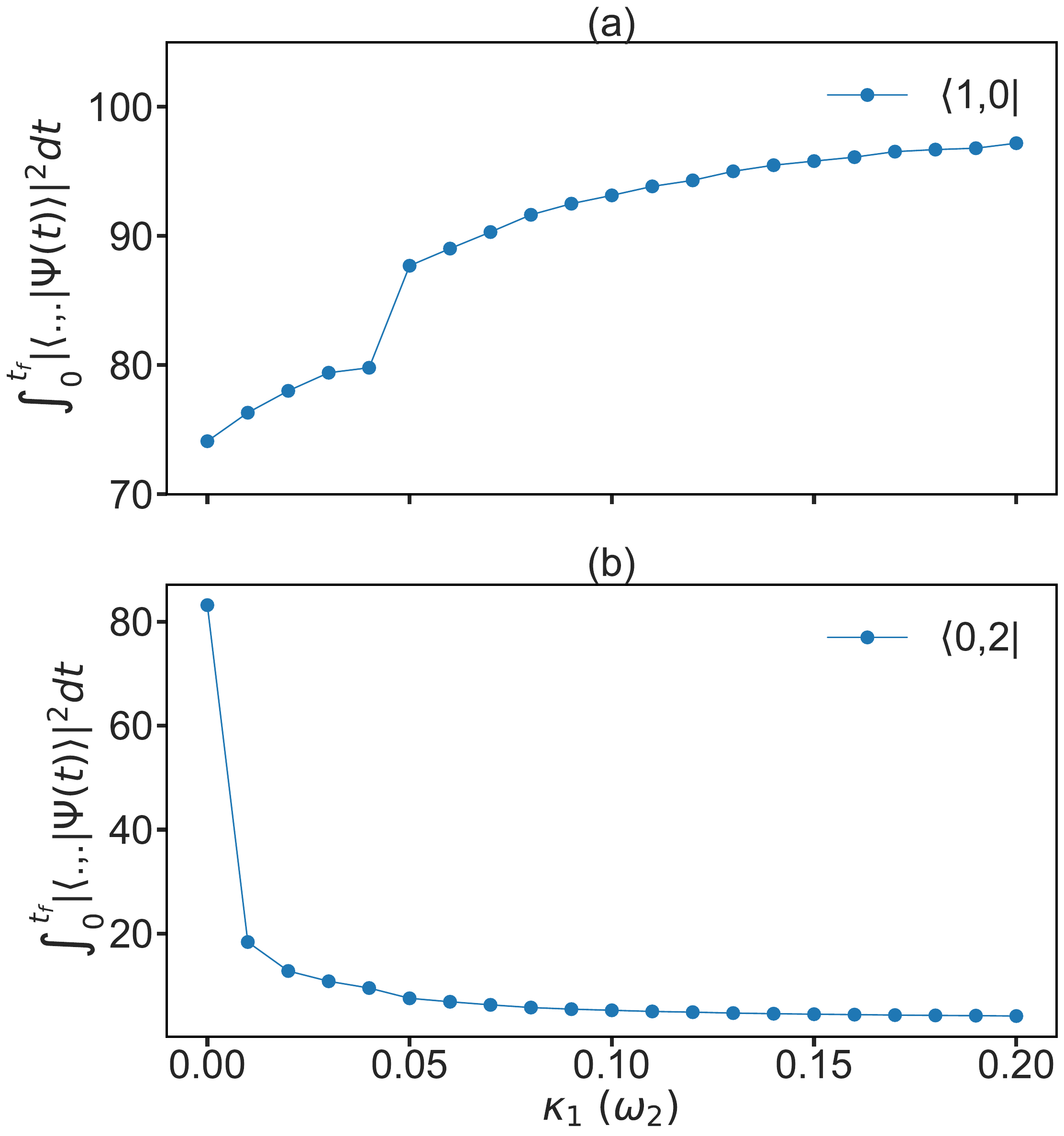}
        \caption{Laser-induced decoupling of the Fermi-resonance pair:  Integrated population of states $|1,0\rangle$ (a) and $|0,2\rangle$ (b).}\label{fig8}
\end{figure}

\begin{figure}[htb]
\centering
        \centering
        \includegraphics[width=0.8\columnwidth]{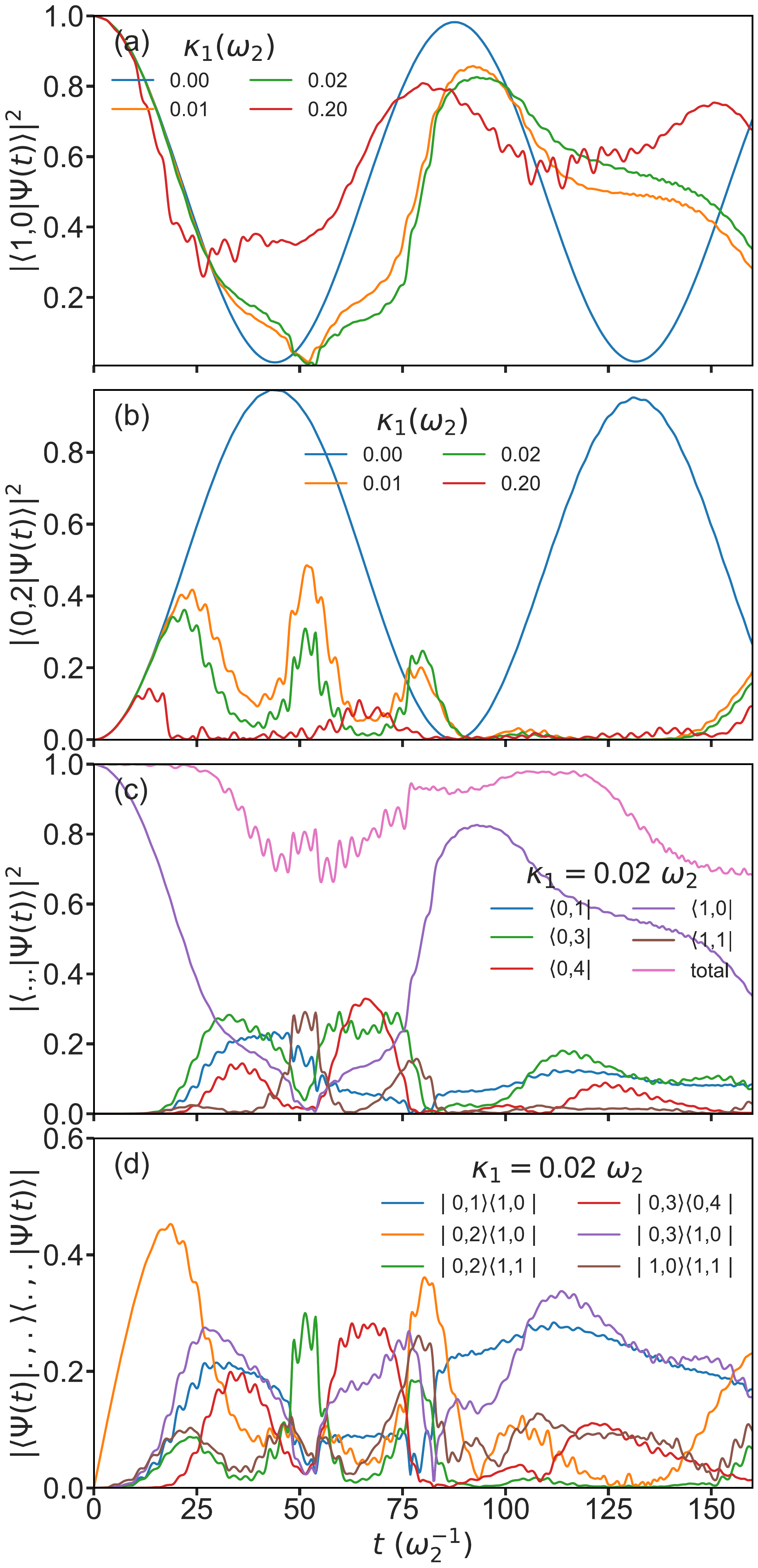}
        \caption{Laser-induced decoupling of the Fermi-resonance pair: Population dynamics of state (a) $|1,0\rangle$ and (b) $|0,2\rangle$ for different values of $\kappa_1$. For $\kappa_1=0.02 ~\omega_2$ the population of the most relevant states are shown in (c) and the most relevant coherences are shown in (d).}\label{fig9}
\end{figure}

\begin{figure*}[htb]
\centering
        \centering
        \includegraphics[width=1.5\columnwidth]{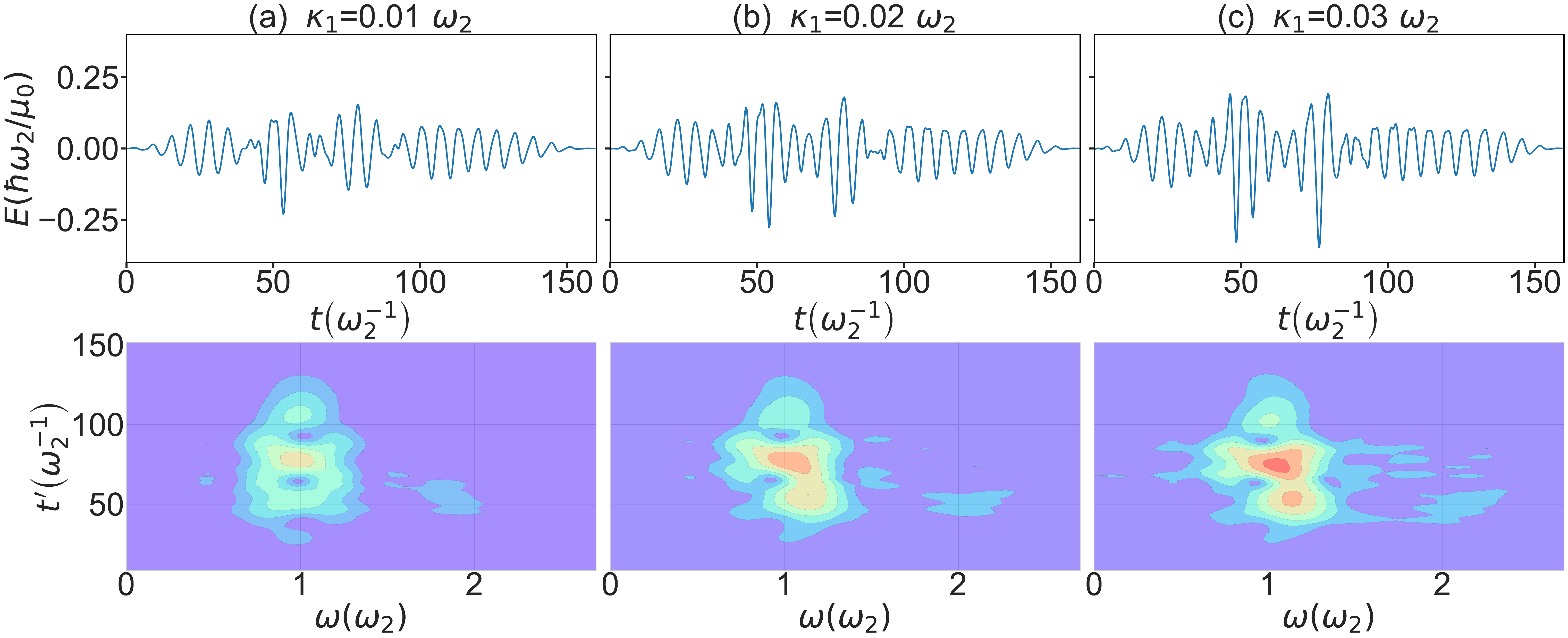}
        \caption{Laser-induced decoupling of the Fermi-resonance pair: Optimal field (first row) and time dependent power spectrum (second row) for different values of $\kappa_1$.}\label{fig10}
\end{figure*}

\subsection{Laser-induced decoupling of the Fermi-resonance pair}\label{sec:example3}

In the previous sections we have demonstrated that it is possible to excite state $|1,0\rangle$, keeping the population of $|0,2\rangle$ on a negligible level. The question that we want to answer in this section is whether it is possible to maintain the population of $|1,0\rangle$ once it has been prepared, or, in other words, can $|1,0\rangle$ be transiently decoupled from $|0,2\rangle$.

Realization of this goal is an example for using the running cost functional ${\mathcal R}_2[\bm a(t)]$, whose integral will be maximized/minimized for the state $|1,0\rangle$/$|0,2\rangle$. Thus, we have $|\phi_1\rangle=|0,2\rangle$, $|\phi_2\rangle=|1,0\rangle$, and  $\kappa_2=-\kappa_1$, which has been ranged just as in the previous sections. \ok{ It turned that the optimal field leads to a population of rather high energetic states. In terms of IVR this is, of course, not a favorable situation. To cope with it, high energetic states have been included into the running costs such as to minimize their popultion. To this end we have used $|\phi_3\rangle=|16\rangle$, $|\phi_4\rangle=|17\rangle$, $\ldots$,  $|\phi_{23}\rangle=|36\rangle$ and $\kappa_3=\kappa_4=...=\kappa_{23}=\kappa_{1}$.}

Figure \ref{fig8} shows that this task can in principle be accomplished. While the integral population of state $|1,0\rangle$ rise above the level of the $\kappa_1=0$ case, the respective population of state $|0,2\rangle$ drops to a low level already for small $\kappa_1$. Clearly, the effect on $|0,2\rangle$ is more pronounced than that on $|1,0\rangle$. As we will see this implies that during the considered time interval the field will couple  $|1,0\rangle$ to states other than $|0,2\rangle$.

Figures \ref{fig9}  and \ref{fig10} show state populations for particular values of $\kappa_1$ and an analysis of the control field, respectively. In case that $\kappa_1=0$ complete population exchange is observed as expected for a coupled two-level system with degenerate levels~\cite{may11}. Upon increasing the penalty one still observes the initial drop in population for state $|1,0\rangle$ as well as the subsequent rise. The following decrease of population is suppressed in case of weak penalties. \textcolor{red}{With increasing penalty ($\kappa_1=0.2\,\omega_2$) the initial drop is not that pronounced, which holds for the subsequent oscillation as well.} Interestingly, panel (b) shows that the zero-order dynamics of $|0,2\rangle$ is ``disrupted'' several times by the laser field. According to panel (c) this proceeds by populating e.g., states $|0,n_2\rangle$ with $n_2=1,3,4$. From the coherences shown in Fig. \ref{fig9}(d) we notice a large element $\rho_{1,0;0,2}$ following the population of state $|0,2\rangle$ in panel (b) and thus giving the Fermi-resonance mediated population exchange. The other coherence, such as  $\rho_{1,0;0,3}$, are field-mediated and serve the redirection of population flow from $|0,1\rangle$ to higher overtones and combination transitions.

The optimal fields corresponding to some of the values of $\kappa_1$  are shown in Fig.~\ref{fig10}. It can be seen the fields are dominated by components at $\omega_2$, i.e. the control mechanism is only marginally involving the ground state. Compared to the other cases, we have a  broader and more structured intensity distribution around $\omega_2$ due to the deviation of the energy spectrum from the harmonic oscillator case, see Fig. \ref{fig:eigenenergies}. The intensity of these fields are also significantly higher.

\section{Summary}
We have presented the first application of DOCT to the control of the dynamics of a quantum system in a finite dimensional Hilbert space. DOCT follows the first discretize and then optimize paradigm. This has the advantage of greater flexibility as compared to standard indirect OCT. The control functional including constraints can be defined and if necessary modified without the need for tedious derivation of working equations using functional calculus. \ok{The disadvantage is that the solution of the nonlinear optimization task using the current implementation is limited to about 80 dynamical variables for problems alike the present one. In addition there is a constraint with respect to the temporal discretization to about 1000 nodes. This limits the overall time interval depending on the rate at which, e.g., the control field changes. It should be stressed, however, that we did not push the numerical limits yet, i.e. all calculations have been performed on a single node of regular hardware.}

For the purpose of illustration DOCT has been applied to control the dynamics of a Fermi-resonance, which is a typical example of anharmonic coupling in vibrational system such as hydrogen bonds. It is well-know that the Fermi-resonance interaction provides the gateway for rapid IVR of the A-H stretching vibration via the A-H bending overtone \cite{heyne04_6083}. To explore possible control strategies avoiding this IVR pathway, three scenarios have been discussed. 
Thereby, we explored the fact that  contrary to the method presented in Ref. \cite{palao08_063412}, in DOCT there is no need to express the state-dependent constraints in terms of maximizing the  population in the allowed space. Instead one  can select the signs of the penalties individually and thus have full control of the populations of the respective subspaces.

First, state-selective optimization of the Fermi-resonance pair population has been performed, i.e. maximization/minimization of population of the fundamental/overtone transitions. While such an optimization is possible, it came at the expense of populating higher overtone and combination states. Therefore, in a second optimization such populations were suppressed. Finally, we asked the question whether an already prepared fundamental excitation can be transiently decoupled from the overtone state. Indeed this was possible, but only at the price of populating higher excited states. Analyzing the optimal fields and the dynamics of the system's density matrix it could be shown that in all cases not only direct ladder-climbing type of transitions play a role, but there is a substantial contribution of an indirect pathway starting from the fundamental excitation of the bending vibration and involving its deexcitation simultaneously to the excitation of the stretching transition. 

Of course, these examples are oversimplifications of the situation, e.g., in real hydrogen bonds, which involves coupling to further intramolecular vibrations as well as to the surrounding solvent~\cite{kuhn02_7671,heyne04_6083}. In cases where a mapping to a system-bath Hamiltonian with perturbative treatment of the system-bath interaction is possible~\cite{may11}, the dynamics could be described by a quantum master equation for the reduced density matrix. Although one has to consider that this increases the number of dynamical variables there is no principal reason why DOCT could not be applied to dissipative dynamics.
\section*{Acknowledgments}
The authors thank the Deutsche Forschungsgemeinschaft (DFG) for financial support through the project Ku952/10-1.

%

\begin{thebibliography}{10}
\expandafter\ifx\csname url\endcsname\relax
  \def\url#1{\texttt{#1}}\fi
\expandafter\ifx\csname urlprefix\endcsname\relax\def\urlprefix{URL }\fi
\expandafter\ifx\csname href\endcsname\relax
  \def\href#1#2{#2} \def\path#1{#1}\fi

\bibitem{keefer18_2279}
D.~Keefer, R.~{de Vivie-Riedle}, Pathways to {{New Applications}} for {{Quantum
  Control}}, Acc. Chem. Res. 51 (2018) 2279--2286.

\bibitem{judson92_1500}
R.~S. Judson, H.~Rabitz, Teaching lasers to control molecules, Phys. Rev. Lett.
  68 (1992) 1500.

\bibitem{assion98_5}
A.~Assion, T.~Baumert, M.~Bergt, T.~Brixner, B.~Kiefer, V.~Seyfried,
  M.~Strehle, G.~Gerber, Control of {Chemical} {Reactions} by
  {Feedback}-{Optimized} {Phase}-{Shaped} {Femtosecond} {Laser} {Pulses},
  Science 282 (1998) 919.

\bibitem{brixner03_418}
T.~Brixner, G.~Gerber, Quantum {{Control}} of {{Gas}}-{{Phase}} and
  {{Liquid}}-{{Phase Femtochemistry}}, ChemPhysChem 4 (2003) 418.

\bibitem{peirce88_4950}
A.~Peirce, M.~Dahleh, H.~Rabitz, Optimal control of quantum mechanical systems:
  {{Existence}}, numerical approximations, and applications, Phys. Rev. A 37
  (1988) 4950.

\bibitem{kosloff89_201}
R.~Kosloff, S.~Rice, P.~Gaspard, S.~Tersigni, D.~Tannor, Wavepacket dancing:
  {{Achieving}} chemical selectivity by shaping light pulses, Chem. Phys. 139
  (1989) 201--220.

\bibitem{brif10_075008}
C.~Brif, R.~Chakrabarti, H.~Rabitz, Control of quantum phenomena: Past, present
  and future, New J. Phys. 12 (2010) 075008.

\bibitem{worth13_113}
G.~A. Worth, G.~W. Richings, Optimal control by computer, Annu. Rep. Prog.
  Chem., Sect. C: Phys. Chem. 109 (2013) 113.

\bibitem{engel09_29}
V.~Engel, C.~Meier, D.~J. Tannor, {Local Control Theory: Recent Applications to
  Energy and Particle Transfer Processes in Molecules}, Adv. Chem. Phys. 141
  (2009) 29--101.

\bibitem{gollub08_073002}
C.~Gollub, M.~Kowalewski, R.~d. Vivie-Riedle, {Monotonic Convergent Optimal
  Control Theory with Strict Limitations on the Spectrum of Optimized Laser
  Fields}, Phys. Rev. Lett. 101 (2008) 073002.

\bibitem{devivie07_5082}
R.~{de Vivie-Riedle}, U.~Troppmann, {Femtosecond Lasers for Quantum Information
  Technology}, Chem. Rev. 107 (2007) 5082--5100.

\bibitem{ohtsuki01_8867}
Y.~Ohtsuki, K.~Nakagami, Y.~Fujimura, W.~Zhu, H.~Rabitz, {Quantum optimal
  control of multiple targets: Development of a monotonically convergent
  algorithm and application to intramolecular vibrational energy redistribution
  control}, J. Chem. Phys. 114~(20) (2001) 8867--8876.

\bibitem{zhao00_4882}
Y.~Zhao, O.~K{\"u}hn, Competitive {{Local Laser Control}} of
  {{Photodissociation Reaction HCo}}({{CO}}){\textsubscript{4}} to
  {{HCo}}({{CO}}){\textsubscript{3}}+ {{CO}} in {{Electronic Ground State}}, J.
  Phys. Chem. A 104 (2000) 4882--4888.

\bibitem{tremblay08_194709}
J.~C. Tremblay, S.~Beyvers, P.~Saalfrank, {Selective excitation of coupled CO
  vibrations on a dissipative Cu(100) surface by shaped infrared laser pulses},
  J. Chem. Phys. 128 (2008) 194709.

\bibitem{palao08_063412}
J.~P. Palao, R.~Kosloff, C.~P. Koch, {Protecting coherence in optimal control
  theory: State-dependent constraint approach}, Phys. Rev. A 77 (2008) 063412.

\bibitem{kumar13_326}
P.~Kumar, S.~A. Malinovskaya, I.~R. Sola, V.~S. Malinovsky, {Selective creation
  of maximum coherence in multi-level $\Lambda$ system}, Mol. Phys. 112 (2013)
  326--331.

\bibitem{ndong14_857}
M.~Ndong, C.~Koch, D.~Sugny, {Time optimization and state-dependent constraints
  in the quantum optimal control of molecular orientation}, J. Mod. Opt. 61
  (2014) 857--863.

\bibitem{lapert12_033406}
M.~Lapert, J.~Salomon, D.~Sugny, {Time-optimal monotonically convergent
  algorithm with an application to the control of spin systems}, Phys. Rev. A
  85 (2012) 033406.

\bibitem{malinovsky97_67}
V.~S. Malinovsky, C.~Meier, D.~J. Tannor, {Optical paralysis in electronically
  congested systems: application to large-amplitude vibrational motion of
  ground state Na$_2$}, Chem. Phys. 221 (1997) 67--76.

\bibitem{reich_2012_Monotonically}
D.~M. Reich, M.~Ndong, C.~P. Koch, Monotonically convergent optimization in
  quantum control using {{Krotov}}'s method, J. Chem. Phys. 136 (2012) 104103.

\bibitem{serban_2005_Optimal}
I.~Serban, J.~Werschnik, E.~K.~U. Gross, Optimal control of time-dependent
  targets, Phys. Rev. A 71 (2005) 053810.

\bibitem{palao_2003_Optimal}
J.~P. Palao, R.~Kosloff, Optimal control theory for unitary transformations,
  Phys. Rev. A 68 (2003) 062308.

\bibitem{betts10_}
J.~T. Betts, Practical {{Methods}} for {{Optimal Control}} and {{Estimation
  Using Nonlinear Programming}}, Advances in {{Design}} and {{Control}},
  {Society for Industrial and Applied Mathematics}, 2010.

\bibitem{pardo16_946}
D.~Pardo, L.~Moller, M.~Neunert, A.~W. Winkler, J.~Buchli, Evaluating {{Direct
  Transcription}} and {{Nonlinear Optimization Methods}} for {{Robot Motion
  Planning}}, IEEE Robot. Autom. Lett. 1 (2016) 946--953.

\bibitem{chen-charpentier20_112983}
B.~M. {Chen-Charpentier}, M.~Jackson, Direct and indirect optimal control
  applied to plant virus propagation with seasonality and delays, J. Comput.
  Appl. Math. 380 (2020) 112983.

\bibitem{ramos21_615168}
A.~R.~R. Ramos, O.~K{\"u}hn, {Direct Optimal Control Approach to Laser-Driven
  Quantum Particle Dynamics}, Front. Phys. 9 (2021) 615168.

\bibitem{henri-rousseau02_241}
O.~Henri-Rousseau, P.~Blaise, D.~Chamma, Infrared lineshapes of weak hydrogen
  bonds: recent quantum developments, Adv. Chem. Phys. 121 (2002) 241.

\bibitem{kuhn02_7671}
O.~K\"uhn, Dissipative {Laser}-{Driven} {Hydrogen}-{Bond} {Dynamics} in
  {Deuterated} o-{Phthalic} {Acid} {Monomethylester}, J. Phys. Chem. A 106
  (2002) 7671--7679.

\bibitem{heyne04_6083}
K.~Heyne, E.~T.~J. Nibbering, T.~Elsaesser, M.~Petkovi{\'c}, O.~K{\"u}hn,
  {Cascaded Energy Redistribution upon O-H Stretching Excitation in an
  Intramolecular Hydrogen Bond}, J. Phys. Chem. A 108 (2004) 6083--6086.

\bibitem{abdel-latif_2011_Carbonyl}
M.~Abdel-Latif, O.~Kühn, Carbonyl vibrational wave packet circulation in
  {{Mn}}{\textsubscript{2}}({{CO}}){\textsubscript{10}} driven by ultrashort
  polarized laser pulses, J. Chem. Phys. 135 (2011) 084314.

\bibitem{lisaj_2014_Laserdriven}
M.~Lisaj, O.~Kühn, Laser-driven localization of collective {{CO}} vibrations
  in metal-carbonyl complexes, J. Chem. Phys. 141 (2014) 204303.

\bibitem{santos18_2213}
L.~Santos, M.~Herman, M.~Desouter-Lecomte, N.~Vaeck, Rovibrational laser
  control targeting a dark state in acetylene. simulation in the {N$_s$ = 1,
  N$_r$ = 5 polyad}, Mol. Phys. 116 (2018) 2213--2225.

\bibitem{werschnik07_R175}
J.~Werschnik, E.~K.~U. Gross, Quantum optimal control theory, J. Phys. B: At.
  Mol. Opt. Phys. 40 (2007) R175--R211.

\bibitem{worth10_15570}
G.~A. Worth, C.~S. Sanz, Guiding the time-evolution of a molecule: Optical
  control by computer, Phys. Chem. Chem. Phys. 12 (2010) 15570.

\bibitem{becerra10_1391}
V.~M. Becerra, Solving complex optimal control problems at no cost with
  {{PSOPT}}, in: 2010 {{IEEE International Symposium}} on {{Computer}}-{{Aided
  Control System Design}}, {Yokohama, Japan}, 2010, pp. 1391--1396.

\bibitem{sundermann00_1896}
K.~Sundermann, R.~de~{Vivie-Riedle}, Extensions to quantum optimal control
  algorithms and applications to special problems in state selective molecular
  dynamics, J. Chem. Phys. 110 (2000) 1896.

\bibitem{doslic06_013402}
N.~Do{\v s}li{\'c}, Generalization of the {{Rabi}} population inversion
  dynamics in the sub-one-cycle pulse limit, Phys. Rev. A 74 (2006) 013402.

\bibitem{may11}
V.~May, O.~K\"uhn, {Charge and Energy Transfer Dynamics in Molecular Systems,
  3rd revised and enlarged edition}, Wiley-VCH, Weinheim, 2011.

\end{thebibliography}
%
%
%
%

\end{document}